# Self-Resetting Soft Ring Enables Autonomous and Continuous Leaping under Uniform Light

Fangjie Qi[1], Caizhi Zhou[1], Haitao Qing[1], Haoze Sun[1], Jie Yin[1*]

[1]Department of Mechanical and Aerospace Engineering, North Carolina State University
Raleigh, NC, 27695 USA

* Email: jyin8@ncsu.edu

## Abstract

Jumping is an efficient locomotion strategy to traverse cluttered, uneven, or unstable environments in nature, yet replicating continuous, autonomous leaping in soft robots remains challenging due to limited energy storage and reliance on human intervention or latches. Here, we report a millimeter-scale, self-resetting soft ring that achieves repeated vertical and stable horizontal leaps under uniform infrared illumination without external control. The ring-shaped liquid crystal elastomer body twists to store elastic energy, which is suddenly released when a rigid tail strikes the ground, propelling the robot. During the airborne phase, the twisted body autonomously untwists, resetting for the next cycle. By tuning geometric asymmetry and the center of mass, the robot transitions between crawling, directional leaping, and vertical jumping. Optimized configurations yield vertical jumps exceeding 80 body heights and directional horizontal leaps over 3 body lengths. Beyond controlled motion on flat ground, the robot demonstrates resilient and robust locomotion across slopes, parallel hurdles, and diverse cluttered natural terrains including grass, wet sand, and mulch. This work establishes a new paradigm of twisting-enabled, photothermally powered soft robots capable of autonomous, continuous leaping, with potential applications in environmental navigation, swarm robotics, and unstructured terrain navigation.

# INTRODUCTION

Jumping is a powerful locomotion strategy in nature, enabling organisms to traverse long distances rapidly and navigate complex terrains. The underlying principle is universal: elastic energy is stored and suddenly released to generate a power-amplified burst of motion. Energy can be stored through bending, stretching, or twisting[1-4], and released through distinct mechanisms—snapping instabilities[1,2,4-9], latching[2,3,10], and explosion[11,12] enable repeated cycles[1,3-9,11], whereas fracturing produces single-use launches[13]. In animals such as fleas, trap-jaw ants, and springtails[14], repetitive leaping is sustained by coupling snap-through structures with muscular resetting or anatomical latches[15,16]. By contrast, plants lack muscles and most often achieve single-shot dispersal, as in squirting cucumbers or legume pods that explosively fracture to scatter seeds in response to perturbation or humidity[13,17]. A few exceptions, including wild oat awns[18] and horsetail spores[19], exploit reversible snapping driven by cyclic humidity changes, achieving limited repeatability[18,19]. Thus, while animals rely on active muscular or latch-based reloading, plants achieve passive resetting via environmentally driven material responses.

Inspired by these strategies, researchers have developed soft jumping devices or robots using stimuli-responsive materials such as liquid crystal elastomers (LCEs)[20-24], hydrogels[13,25], electroactive polymers[26], and magneto-elastomers[7,27]. Yet a central challenge persists achieving autonomous, continuous, and adaptive leaping without latches, active stimulus modulation, or human intervention. Analogous to the origami "paper frog", which jumps once before requiring manual resetting, most soft jumpers are effectively single-shot devices. It requires manual resetting or dynamic modulation of external stimuli (e.g., toggling electric, magnetic, light, or thermal inputs on and off) to operate repetitively in either tethered or untethered modes[3,7,13,21,23,26,28-31]. Recent advances—such as evaporating elastomeric disks that self-snap to bounce a few times[32], or combustion-powered jumpers[12] that leap forcefully but only for limited cycles—demonstrate novel strategies, yet none achieve sustained, untethered autonomy.

Four obstacles underpin this challenge. First, energy resetting under a constant field is inherently difficult: most engineered jumpers require time-varying stimuli or manual flipping between bistable states to recover shape[3,7,13,21,23,26,28-30], complicating both actuation and control. Second, compact energy storage and recycling remain inefficient—bending, stretching, or combustion-based mechanisms demand large shape changes and external resetting, limiting leaping repeatability at small scales. Third, aerial posture regulation is critical: body orientation during flight dictates landing state, and without stable recovery, sustained and directional leaping is impossible. Fourth, pre-jump posture adjustment is often missing: whereas animals shift their center of mass to steer trajectory, soft robots typically lack such self-regulation. Overcoming these barriers is crucial for achieving robust and continuous soft jumping.

Here, we report a self-resetting directional soft ring jumper that establishes a new design paradigm (**Fig. 1a**): cyclic elastic energy storage and release without latches or active control under constant fields. Under uniform infrared illumination, the liquid crystal elastomer ring twists to store torsional energy, which is suddenly released via self-snapping when its rigid tail strikes the ground, launching the body into the air. Twisting serves as a compact, efficient torsional spring, storing energy more effectively than bending or stretching. During flight, the structure autonomously untwists and resets via geometric inversion symmetry, enabling continuous relaunching (**Fig. 1b-**

**1d**). By tuning geometric asymmetry and mass distribution, we realize programmable transitions among crawling, vertical jumping, and directional leaping with remarkable landing stability. Beyond sustained leaping, the soft ring demonstrates robust, self-adaptive locomotion across slopes, obstacles, and natural terrains, providing a generalizable framework for autonomous, sustained soft machines powered by uniform energy fields.

**RESULTS**

**Continuous leaping of the soft ring**

Inspired by springtails that snap appendages to leap, we designed a hybrid millimeter-size "tailed" LCE ring (body length 20 mm and body height 1 mm) by joining the ends of a soft LCE rod with a short aluminum tube, creating a sharp binding angle $\beta$ that breaks circular symmetry (**Fig. 1a**). The structure integrates a soft, deformable LCE segment as the main body with a rigid aluminum sheet-reinforced segment as a triangular tail (Methods). The tail's rigidity and sharp angle raise the energy barrier for release, allowing higher elastic energy accumulation and enhancing the striking force against the ground (**Fig. 1b**).

Under uniform infrared (IR) illumination, the ring body undergoes photo-thermal heating-induced anisotropic shrinking, flipping outside-in due to temperature gradients [33-35] and progressively storing strain energy through torsional twisting (**Fig. 1c**) (Methods). The rigid tail, resisting twist, bends and rotates upward toward the ring's center. When the triangular tail approaches a near-parallel orientation with the ground, it snaps inward within ~30 ms, rapidly releasing stored energy to strike the surface and launch the body into the air with a launching angle $\alpha \sim 47°$ ($\beta = 100°$, light intensity $w = 0.6$ W·cm$^{-2}$, **Fig. 1c**). The leaping process is reproduced by the corresponding finite element method simulation.

Crucially, once airborne, the twisted LCE segment relaxes back to its untwisted but inverted state (**Fig. 1d**). Because of the ring's inversion symmetry, this relaxation resets the geometry automatically before landing. It achieves a horizontal jumping distance of ~ 34 mm, i.e., ~ 1.7 body lengths (BLs). Under continuous IR illumination, the cycle repeats: on-ground twisting stores energy, followed by in-air untwisting that restores the initial configuration (**Fig. 1d**). This closed, limit-cycle process enables fully autonomous and repetitive leaping (over 500 jumps) without latches, external modulation, or manual resetting.

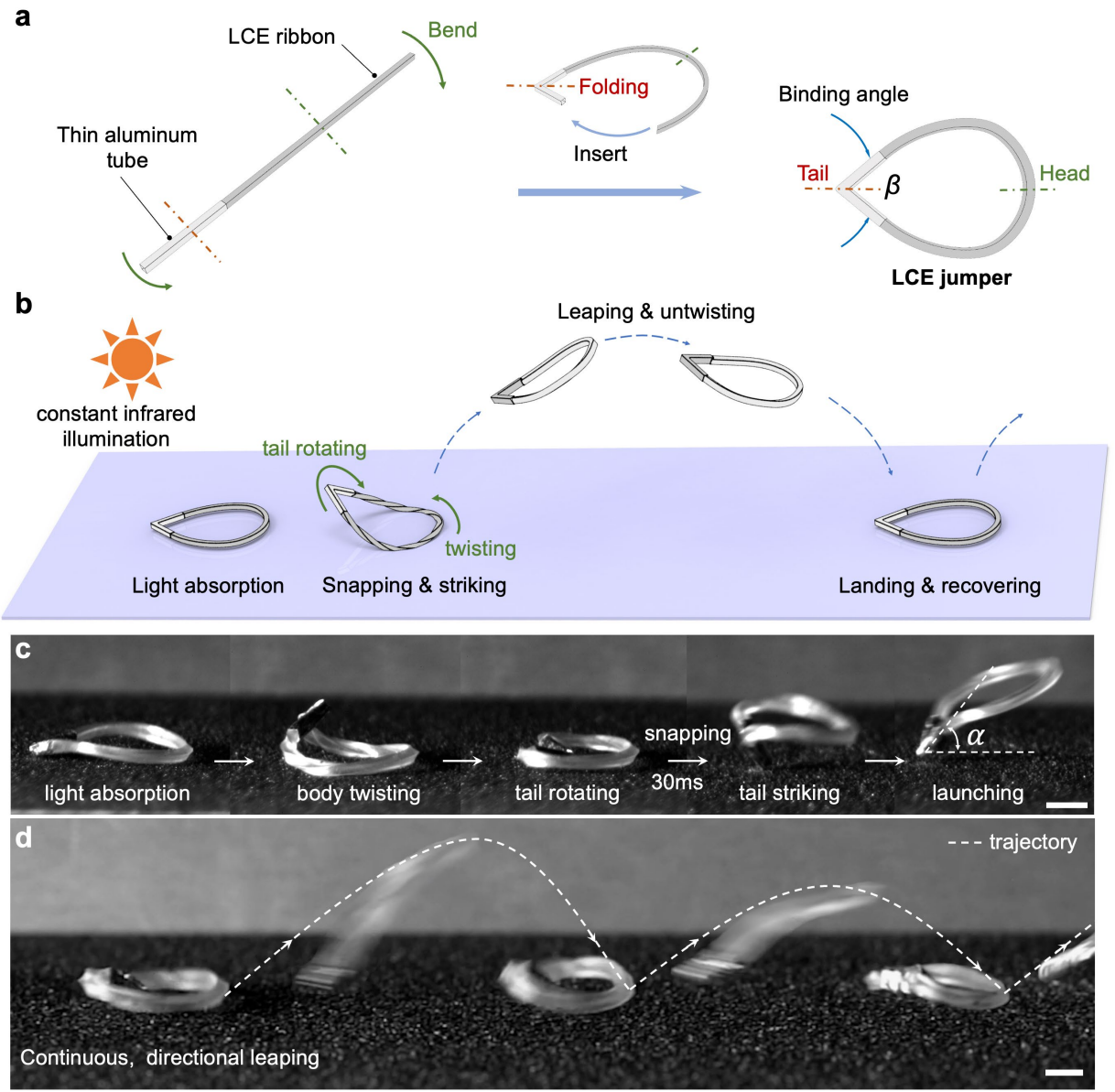

**Fig. 1 Self-resetting soft ring for continuous and autonomous leaping. a**, Fabrication of the jumping ring by bending a liquid crystal elastomer (LCE) ribbon and binding its ends with a thin braided aluminum tube to form a stiff triangular tail with binding angle $β$. **b**, Schematics of self-resetting and continuous leaping under constant infrared illumination through cyclic torsional energy storage, release, launch, airborne untwisting, landing, and recovery. **c**, High-speed image sequence of the launch process showing body twisting and tail rotation for elastic energy storage, followed by tail snapping to strike the ground and release the stored energy. **d**, Continuous, directional leaping tracked under constant illumination. Dashed lines mark head trajectories. Scale bars: 5 mm

## Locomotion phase diagram

The interplay between geometry and energy input dictates the ring's locomotion modes. By varying the binding angle $\beta$, which sets the energy barrier, and the light intensity $w$, which controls actuation power, we mapped a phase diagram spanning vertical jumping, horizontal leaping, and crawling (**Fig. 2a-2b**).

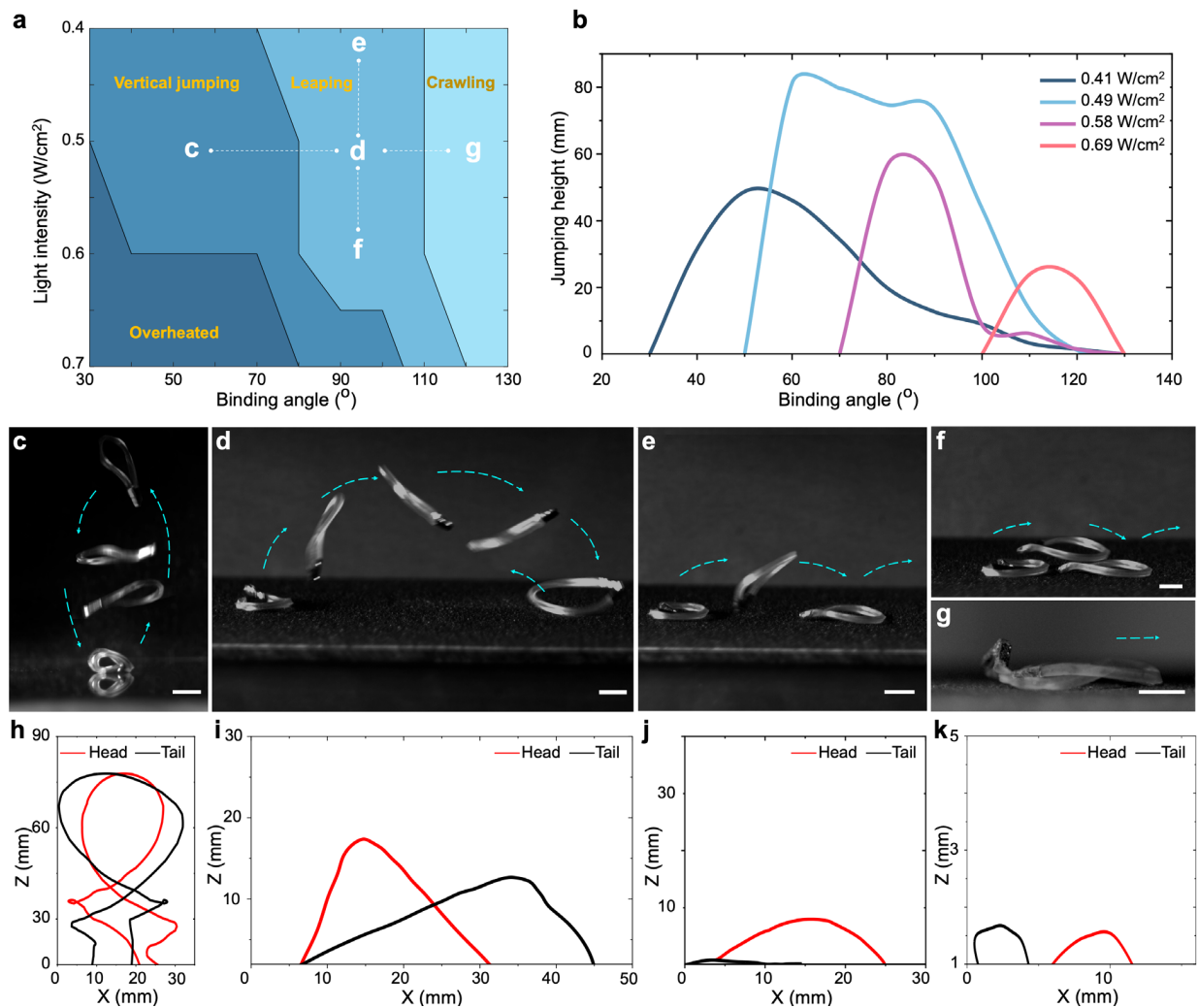

**Fig. 2 Locomotion phase diagram governed by geometry and illumination. a**, Phase diagram showing transitions among continuous vertical jumping, horizontal leaping, and crawling as functions of binding angle $\beta$ and light intensity $w$. **b**, Jumping height as a function of $\beta$ under varying $w$. **c**, Sharp binding angles ($\beta = 60°$, $w = 0.49$ W·cm$^{-2}$) produce vertical jumps via a self-locking mechanism. **d–f**, Intermediate angles ($\beta = 95°$) generate forward leaping: backflipped leaping under intermediate $w = 0.49$ W·cm$^{-2}$ (d), stable forward leaps under lower $w = 0.41$ W·cm$^{-2}$ (e), and shortened low-angle leaps at higher $w = 0.58$ W·cm$^{-2}$ (f). **g**, Large binding angles ($\beta = 120°$) produces slow crawling due to low torsional energy. **h–k**, Corresponding head and tail trajectories for the motions shown in c–f. Scale bars, 5 mm.

At low-to-moderate light intensity (0.4–0.6 W·cm$^{-2}$), increasing $\beta$ from 30° to 130° drives a clear transition (**Fig. 2a**): sharp tails (30° ≤ $\beta$ ≤ 80°) enable vertical jumps (**Fig. 2c**), intermediate $\beta$ (80° < $\beta$ < 110°) produce leaping (**Fig. 2d-2f**), and near-circular rings (110° < $\beta$ ≤ 130°) crawl (**Fig. 2g**). For small $\beta$, triangular tails raise the energy barrier, allowing greater torsional energy storage before release. At shaper angles of 30°–80°, the rings execute powerful vertical jumps. At liftoff, the soft body straightens with a launch angle of ~90°, reaching heights above 80 body heights (BHs) at $\beta$ = 60°, often with in-air flips (**Figs. 2c, 2h**). This performance is enabled by a self-locking effect in which the rotating tail briefly presses against the head before ground contact, delaying release for more energy accumulation that amplifies the strike, while the inversion symmetry of the ring resets the body mid-air for repetitive jumping. At large $\beta$ (110°–130°), the rings approach circularity and behave as crawlers: the tail bends inward too easily, lowering the barrier and producing weak strikes that generate only slow crawling with minimal lift. Between these extremes lies a narrow leaping window (80°–110°), where stored energy propels the ring forward rather than straight upward. At $\beta$ ≈ 95°, the ring launches at ~50° with a straight body, producing ~72 mm leaps (~3.6 BLs) but also a 180° backflip that reverses orientation upon landing (**Figs. 2d, 2i**), yielding alternating headings.

Light intensity further modulates leap stability at a given $\beta$ (**Fig. 2a**). For $\beta$ ≈ 95°, reducing $w$ from 0.49 to 0.41 W·cm$^{-2}$ lowers stored energy, suppressing flips, and enables stabilized forward leaps of ~2 BLs at a consistent ~ 53° launch angle (**Fig. 2e**). During liftoff, the head bends forward to maintain heading in flight, with the tail landing first (**Fig. 2j**). At higher intensities (~0.58 W·cm$^{-2}$), thermal softening decreases the launch angle (~ 20°) and shortens leap distances (~1.1 BL) (**Fig. 2f, 2k**). At very high intensities ($w$ > 0.65 W·cm$^{-2}$), shrinkage-induced interlocking suppresses forward motion in rings with intermediate $\beta$ (80°–100°), reducing them to low-amplitude vertical hopping, while rings with small $\beta$ (30°–70°) overheat and become immobile (**Fig. 2a**). Thus, an optimal intermediate intensity ($w$ ≈ 0.49 W·cm$^{-2}$) maximizes elastic energy storage across a broad range of $\beta$ (55°–110°), enabling jumps of 10–80 BHs (10–80 mm). This phase landscape illustrates how geometry and energy input cooperatively govern transitions between crawling, leaping, and vertical jumping.

**Optimizing controllable leaping by shifting the center of mass**

The phase diagram reveals that stabilized leaping occurs only within a narrow design window, as backflips and alternating headings undermine directional control. This instability originates from misalignment between the center of mass (CoM) and thrust at liftoff, which generates torque and induces rotation at high launch angles. Aligning the CoM with thrust is therefore essential for stability.

To broaden the design space and achieve controlled, longer leaps, we introduce a simple modification: attaching a copper-foil tube as a head weight (**Fig. 3a**). We define the weight ratio (wt%) as the added head mass relative to the net ring mass. As wt% increases from 0 to 36%, the CoM shifts progressively forward (**Fig. 3b**, Methods), suppressing flips, converting vertical jumps into forward leaps, and stabilizing trajectories for extended travel (**Fig. 3c**). This strategy echoes natural jumpers such as grasshoppers and fleas, which bias mass forward to stabilize and steer motion[36].

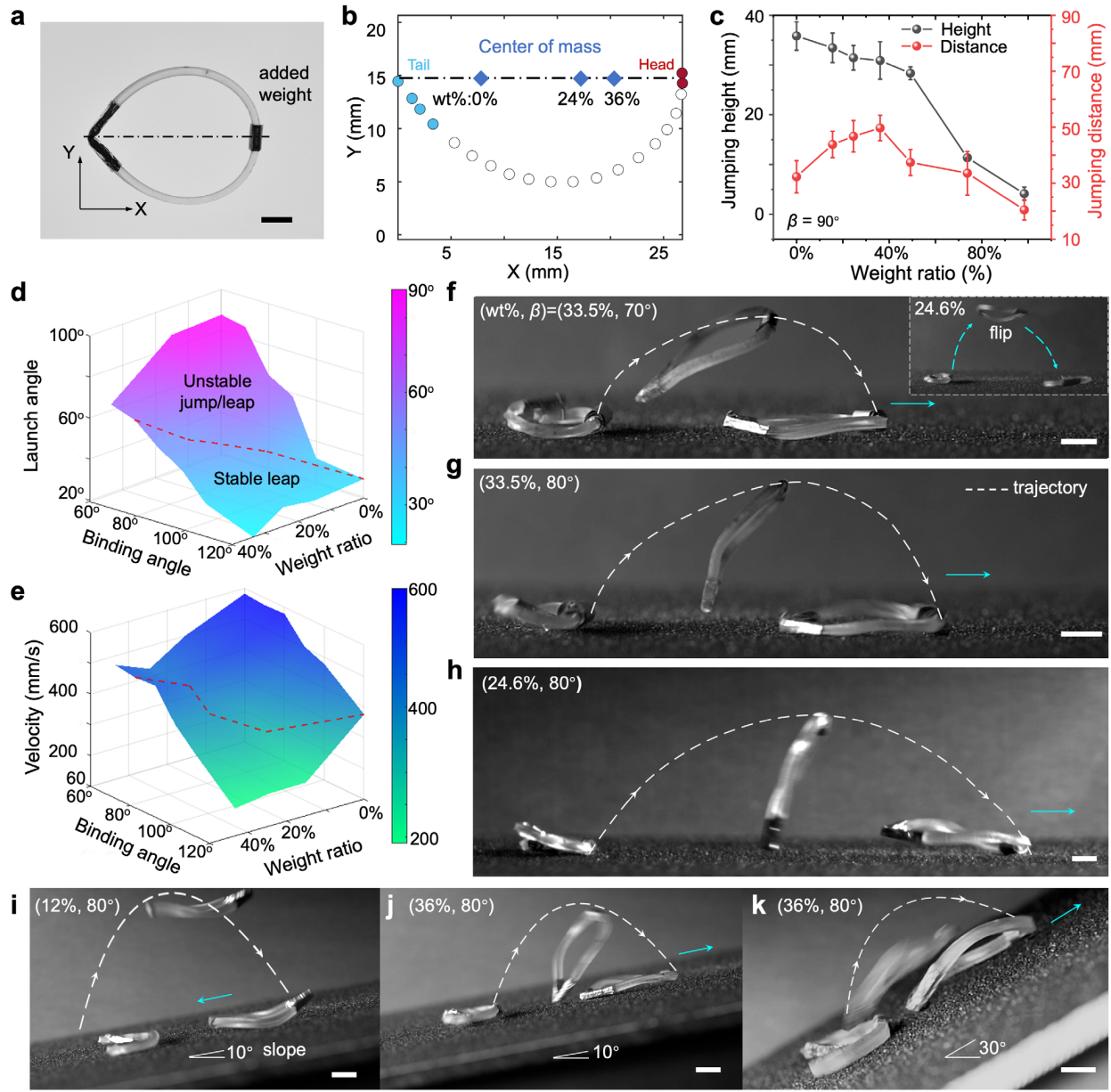

**Fig. 3 Center-of-mass tuning enables robust and directional leaping. a**, A copper-foil tube attached to the head serves as a tunable weight for controlled mass redistribution. **b**, Forward shift of the center of mass with increasing head weight ratio (wt%, 0–36%). Blue, white, and red markers indicate the aluminum-covered tail, pure LCE body, and copper-covered head, respectively. **c**, Jump height and leap distance versus wt% ($\beta = 90°$, $w = 0.49$ W·cm$^{-2}$) reveal an optimal range for maximal horizontal travel. **d–e**, Phase maps of launch angle (**d**) and velocity (**e**) as functions of $\beta$ and wt%, showing a clear transition from unstable flipping to stable, directional leaping (dashed boundary). **f–h**, Representative trajectories illustrating the evolution from backflipping (f, inset) to stable forward leaping (f-g) and optimal, longest leap (~3.1 BLs) (h). **i–k**, Stable leaping on inclined terrains (10° and 30° slopes) achieved through mass tuning. Increasing wt% suppresses backflips and somersaults, enabling reliable uphill locomotion. Dashed lines trace head trajectories. Scale bars, 5 mm.

Theoretically, optimal leaping requires maximizing the horizontal component of translational speed while minimizing rotation. From projectile mechanics, leap distance follows $L = v^2 \sin(2\alpha)/g$, where $v$ is launch speed, $g$ is gravity, and the ideal launch angle is ~45°. Guided by this principle, we tune both binding angle $\beta$ and wt% to maximize speed while achieving forward launch angles of 40°–50° under optimal irradiance ($w \approx 0.49$ W·cm$^{-2}$).

A representative case of a jumping ring with $\beta = 90°$ illustrates the effect of mass redistribution (**Fig. 3c**). As wt% increases from 0 to 100%, it transitions from vertical jumping to stabilized horizontal leaping, where its maximum jumping height decreases monotonically and dramatically when wt% is over 50% due to its increasing weight. However, horizontal leap distance first increases, peaking at ~50 mm (~2.5 body lengths) for wt% ≈36%, before declining. This suggests that appropriate mass redistribution can enhance forward leap distance. Similar trends are observed across other $\beta$.

Maps of launch angle $\alpha$ and speed $v$ as functions of $\beta$ (60°–120°) and wt% (0–33.5%) (**Figs. 3d, 3e**) generalize these behaviors. Small $\beta$ and low wt% produce unstable flips with high $\alpha$ and $v$. Increasing $\beta$ or wt% lowers $\alpha$ by shifting the CoM forward, suppressing flips and enabling directional leaps (**Fig. 3f**). A clear phase boundary emerges: stable leaping requires $\alpha \leq 59°$ and $v \leq 0.46$ m/s. For example, sharp-angle vertical jumpers or unstable back-flippers ($\beta = 60°$–80°) transition into forward leapers ($\alpha$~48°–60°) with added head weight (**Fig. 3f-3h**). Although both $\beta$ and wt% reduce launch speed—via diminished elastic storage or added mass—$\beta$ exerts the stronger effect. The optimal regime occurs at a launch angle of 40°–50° and velocity exceeding 0.4 m/s, achieved at $\beta = 80°$ and wt%=25% (**Fig. 3h**). This configuration yields the longest stabilized leaps of ~75 mm (~3.1 BLs) (**Fig. 3g-3h**), more than twice the distance of electrohydrostatically driven tethered soft jumpers (1.46 BLs) powered by high–energy-density dielectric actuators[3].

Leaping on sloped surfaces presents additional challenges absent on flat ground. At launch, the incline reduces effective thrust and increases slip risk, while at landing, asymmetric impact and downslope gravity bias generate destabilizing torques that induce tumbling and hinder recovery. These effects compromise both efficiency and stability. CoM tuning, however, enables controlled leaping even on inclined terrain. On mild slopes (10°), increasing wt% from 12% to 36% suppresses backflips (**Fig. 3i**) and yields stable uphill leaps of ~ 1.8 BLs (**Fig. 3j**). On steeper slopes (30°), jumpers with low wt% (≤ 12%) backflip at takeoff and somersault upon landing, often falling off the slope. By contrast, a 36% head weight effectively suppresses flips, enabling forward leaping on steep terrain despite minor slip (**Fig. 3k**).

**Unstructured environments with obstacles and natural terrains**

We next investigate continuous leaping in more realistic environments with obstacles and natural terrains.

Obstacle resilience—the ability to sustain locomotion despite collisions—is critical for real-world navigation in cluttered environments. To test this, we place two parallel hurdles (**Fig. 4a-4b**) each 5 mm high (~5 BHs) and spaced 27 mm apart (1.35 BLs). Two distinct outcomes emerge. In the first, the jumper performs clean clearance, leaping ~1.5 BLs to land between the hurdles before vaulting the second (**Fig. 4a**). In the second, collision-assisted clearance turns apparent failure into advantage: when the tail catches on the first hurdle, self-resetting twists the body downward, pressing the bent tail against the hurdle wall to generate extra thrust (**Fig. 4b**). This interaction produces a higher, longer leap (~2.2 BLs), enabling successful traversal of the second hurdle. Even against taller barriers (~26 BHs), jumpers with sharp binding angles manage to vault despite impact. In head-on collisions, the ring sometimes flips but resumes leaping in the opposite direction due to its bottom symmetry, effectively avoiding the obstacle. These results highlight how soft body-obstacle interactions and geometric symmetry confer resilience, enabling robust traversal in cluttered terrains.

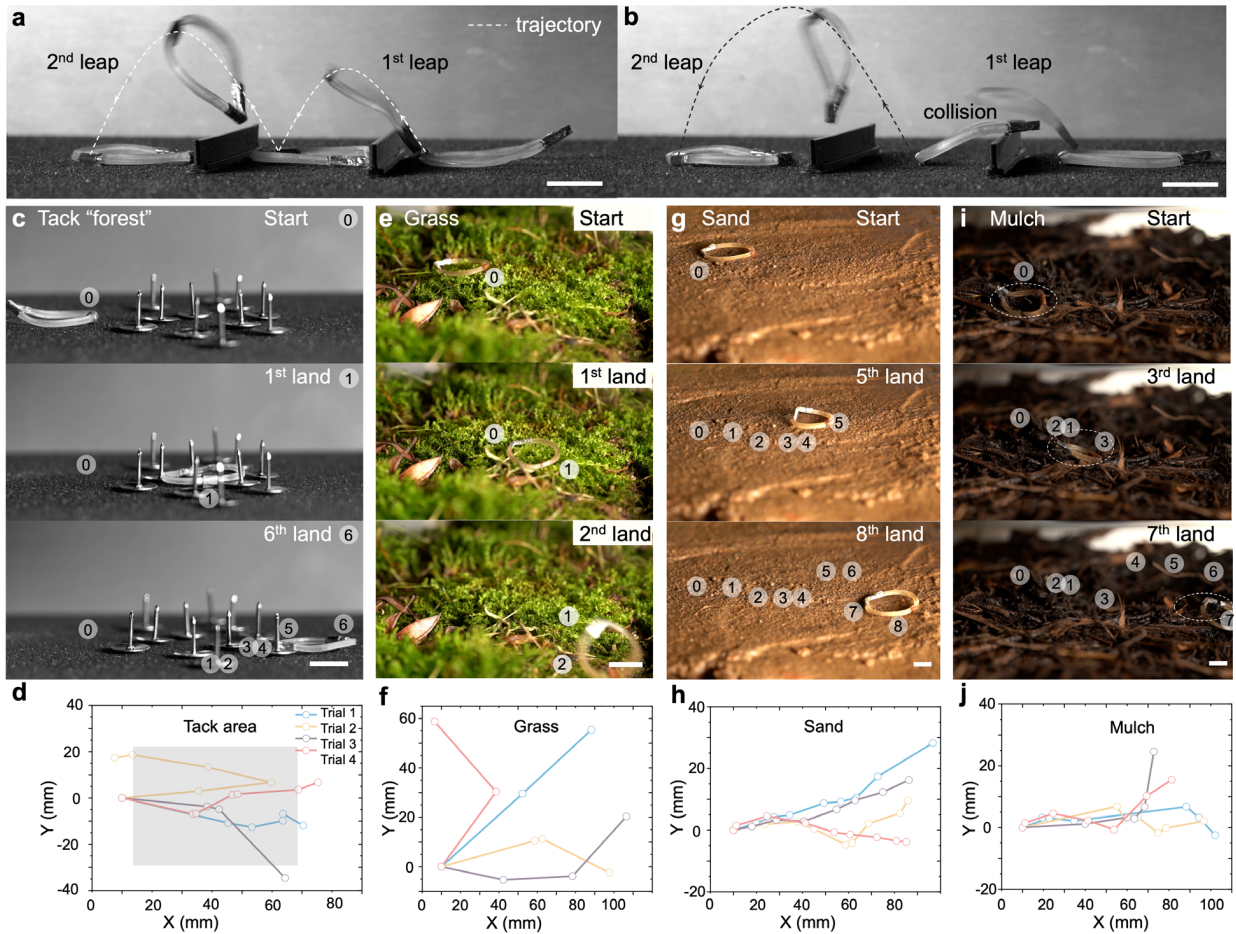

**Fig. 4. Robust continuous leaping across obstacles and natural terrains. a–b**, Obstacle resilience tests over two parallel hurdles (5 BHs high, 1.35 BLs apart). The jumper either clears both hurdles cleanly (**a**) or leverages tail–obstacle interaction for a collision-assisted leap that achieves greater height and distance (**b**). **c–d**, Navigation through randomly scattered tacks simulating a "pillar forest" (5 BHs high). Four representative trajectories of the head are shown in **d**, where each circle marks a landing position. The jumper self-adjusts through successive jumps to escape trapping and resume leaping. **e–j**, Leaping on natural terrains: grass (**e–f**), wet sand (**g–h**), and mulch (**i–j**). Four tracked in-plane trajectories of the head are shown in each case. Grass enables high but erratic jumps, sand provides smoother and more stable motion, and mulch exhibits intermediate behavior.

To further challenge the system, we simulate a "pillar forest" with randomly scattered tacks (5 BHs high; **Fig. 4c-4d**). The jumper occasionally becomes trapped or degrades into crawling, yet through six successive jumps and posture self-adjustments, it can eventually escape and reestablish forward leaping.

We then assess performance on natural terrains including grass, wet sand, and mulch. On grass (**Fig. 4e**), irregular and compliant surfaces induce erratic headings (**Fig. 4f**) but enable consistently high jumps (>4 BHs), enhancing speed despite reduced control. In contrast, motion on wet sand (**Fig. 4g**) requires more steps (8 jumps) and longer travel time, but trajectories are smoother and more predictable (**Fig. 4h**), as the yielding yet solid substrate stabilizes heading. Mulch exhibits intermediate behavior (**Fig. 4i**): irregularity occasionally traps the jumper or induces backward leaps, but overall predictability exceeds that of grass. Trajectory comparisons (**Figs. 4f, h, j**) reveal distinct terrain effects: over 100 mm of travel, sand and mulch support more consistent heading control than grass, with sand offering the most stable trajectories, mulch moderate unpredictability, and grass the fastest but least directed leaps.

Together, these results demonstrate that the self-resetting soft ring sustains robust locomotion across terrains of varying compliance and irregularity, underscoring its potential for navigation in unstructured environments.

**Discussion**
We demonstrate a self-resetting soft ring that achieves sustained, directional leaping under uniform infrared illumination, addressing the long-standing challenge of continuous operation in soft jumpers. The tailed ring couples body twisting and tail bending to cyclically store and release elastic energy, while its inversion symmetry enables automatic resetting upon landing. Simple tuning of mass distribution stabilizes posture and recovery, enabling robust, repeatable, and self-regulate motion across inclined, obstructed, and irregular natural terrains.

Beyond a specific material system, this mechanism reveals a general strategy for autonomous elastic energy recycling in soft active matter. Twisting serves as a compact and efficient mode of elastic energy storage and release—requiring minimal volumetric shape change yet delivering high energy density—making it more effective for cyclic actuation than bending, stretching, or inflation-based mechanisms. This principle is broadly applicable to diverse soft active materials, including photothermal, photoresponsive, magnetic, electroactive, and hydrogel systems, operating under uniform or spatially varying fields such as sunlight[37] or rotating magnetic gradients[7]. The tail, meanwhile, offers a tunable platform for amplifying energy output. Replacing the triangular tail with other thin, flexible sheets of tailored stiffness can enhance bending deformation and boost leap power.

This work establishes a blueprint for autonomous soft machines capable of robust navigation in cluttered, unstructured environments. Such jumpers could be scaled, networked, or deployed as swarms for distributed sensing, monitoring, and exploration in real-world settings where conventional robots fail.

**Acknowledgments**

J. Y. acknowledges the funding support from the National Science Foundation under award number CMMI-2005374, 2369274, and 2445551.

**Author Contributions Statement**

F.Q. and J.Y. designed research. F.Q. fabricated and designed the ring jumpers and characterized their jumping and leaping performances. C.Z. and F.Q. conducted the modeling and finite element simulation. H.Q. and F.Q. conducted the high-speed imaging of the jumping and leaping. F.Q. and H.S. conducted the experimental demonstration in natural terrains. All authors analyzed data. J. Y. supervised the overall research. F. Q. and J. Y. wrote the paper, and all the co-authors revised the paper.

**Competing Interests Statement**

The authors declare no competing interests.

**Data Availability:**
All data generated or analyzed during this study are included in the manuscript and its Supplementary Information.